\documentclass{aa}  

\usepackage{url}
\usepackage{graphicx}
\usepackage{bm}
\usepackage{txfonts}
\begin{document}

   \title{Multi-Scale Deep Learning for Estimating Horizontal Velocity Fields on the Solar Surface}


   \author{Ryohtaroh T. Ishikawa
          \inst{1,2},
          Motoki Nakata\inst{3},
          Yukio Katsukawa\inst{1,2},
          Youhei Masada\inst{4},
          \and
          Tino~L. Riethm{\"u}ller\inst{5}
          }

   \institute{Department of Astronomical Science, School of Physical Sciences, The Gradiate University for Advanced Studies, SOKENDAI,
              2-21-1 Osawa, Mitaka, Tokyo 181-8588, Japan
              \email{ryohtaroh.ishikawa@grad.nao.ac.jp}
         \and
             National Astronomical Observatory of Japan,
             2-21-1 Osawa, Mitaka, Tokyo 181-8588, Japan
         \and
             National Institute for Fusion Science, 322-6 Oroshi-cho, Toki, Gihu 509-5292, Japan
         \and
            Department of Physics and Astronomy, Aichi University of Education, Kariya, Aichi 446-8501, Japan
         \and
            Max-Planck-Institut f\"{u}r Sonnensystemforschung (MPS), Justus-von-Liebig-Weg 3, 37077, G\"{o}ttingen, Germany
             }

   \date{Received ***; accepted ***}

 
  \abstract
   {The dynamics in the photosphere is governed by the multi-scale turbulent convection termed as granulation and supergranulation.
   It is important to derive three-dimensional velocity vectors to understand the nature of the turbulent convection
   and to evaluate the vertical Poynting flux toward the upper atmosphere.
   The line-of-sight component of the velocity can be obtained by observing the Doppler shifts.
   However, it is difficult to obtain the velocity component perpendicular to the line-of-sight,
   which corresponds to the horizontal velocity in disk center observations.}
   {To develop a new method based on a deep neural network that can estimate the horizontal velocity from the spatial and temporal variations
of the intensity and vertical velocity and to suggest a new measure for examining the performance of the method.}
   {We developed a convolutional neural network model with a multi-scale deep learning architecture.
The method consists of multiple convolutional kernels with various sizes of the receptive fields,
and it performs convolution for spatial and temporal axes.
The network is trained with data from three different numerical simulations of turbulent convection.
Furthermore, we introduced a novel coherence spectrum to assess the horizontal velocity fields that were derived at each spatial scale.}
   {The multi-scale deep learning method successfully predicts the horizontal velocities for each convection simulation
in terms of the global-correlation-coefficient, which is often used for evaluating the prediction accuracy of the methods.
The coherence spectrum reveals the strong dependence of the correlation coefficients on the spatial scales.
Although coherence spectra are higher than 0.9 for large-scale structures,
they drastically decrease to less than 0.3 for small-scale structures wherein the global-correlation-coefficient indicates a high value of approximately 0.95.
By comparing the results of the three convection simulations,
we determined that this decrease in the coherence spectrum occurs
around the energy injection scales, which are
characterized by the peak of the power spectra of the vertical velocities.}
   {The accuracy for the small-scale structures is not guaranteed solely by the global-correlation-coefficient.
To improve the accuracy in small-scales, it is important to improve the loss function for enhancing the small-scale structures
and to utilize other physical quantities related to the non-linear cascade of convective eddies as input data.}

   \titlerunning{Multi-Scale Deep Learning for Estimating Horizontal Velocity Fields}
   \authorrunning{R.~T.~Ishikawa et al.}
   \maketitle
%

\section{Introduction}
The dynamics in the solar photosphere are governed by thermally driven convection. This in turn
produces cellular patterns termed as granules that are observed in visible-light continuum images.
The bright areas with hot rising flows are surrounded by darker and cooler intergranular lanes.
A typical granule has a diameter of approximately 1000 km and lasts for approximately 10 min (\citealt{Nordlund09} and references therein).
The turbulent nature of the granular convection inherently creates small-scale flow structures
that are smaller than the typical size of granules
(e.g. \citealt{Matsumoto10}, \citealt{Katsukawa12}, \citealt{Rempel18}).
These types of small-scale flows interact with magnetic fields and can produce the Poyinting flux
upward. This in turn can drive various phenomena, such as explosions \citep{Shibata07,Toriumi17},
jets \citep{Hollweg82,Iijima17},
and heating \citep{vanBallegooijen11,DePontieu12}, in the upper atmosphere, chromosphere, and corona.
Supergranulation is another convective patterns observed on the solar surface,
which is characterized by a horizontal flow fields with the large spatial scale of about 30~Mm \citep{Rieutord10b}
and the typical lifetime of about 1.7 days \citep{Hirzberger08}.
Photospheric magnetic fields are passively advected into the edges of supergranules and form network structures \citep{Gosic14}.
Recent observation found that persistent vortex flows exist at supergranular vertices,
and magnetic flux can be concentrated in the vortices \citep{Requerey18}.

We can obtain the line-of-sight (LOS) component of the flow velocities by a spectroscopic observation via the Doppler effect.
Conversely, to date, there are no direct methods for observing the components perpendicular to the LOS.
These components correspond to the horizontal velocity on the solar surface in disk center observations.
The most commonly used method for estimating the horizontal velocity field is local correlation tracking (LCT; \citealt{November88}).
This method uses two consecutive images and computes the cross-correlation, and thereby detects the motions of granule patterns.
Although the LCT technique can evaluate the horizontal velocity with good accuracy on a larger scale,
its accuracy on a scale smaller than granules is limited
by as much as factor of three \citep{Verma13} or more \citep{Malherbe18}.
The errors are preferentially high in the boundaries between granular cells \citep{Louis15}.
This is mainly due to the fact that the window utilized in the LCT method to compute the cross-correlations blurs velocity fields.
However, the accuracy on a smaller scale is important
for evaluating the interaction between magnetic fields and horizontal flows
because the magnetic fields are often concentrated in small regions in the photosphere (e.g., \citealt{Parnell02}).
	
An alternative approach involves identifying features that are observed as bright points in G-band or continuum images
and obtaining horizontal velocities by tracking them \citep{Berger98,Utz10}.
The method can provide horizontal velocities of small magnetic features in intergranular lanes.
Given that such magnetic features appear associated with strong concentrations of magnetic fields,
we cannot obtain velocity fields and their spatial distribution across the entire areas by using this method.

There is a new method for estimating the horizontal velocity that employs a deep learning approach.
\citet{Asensioramos17} developed a model using a convolutional neural network (DeepVel), which was
trained on a set of velocity fields simulated for the photosphere.
DeepVel can estimate the horizontal velocity at various heights in the solar atmosphere without averaging.
\citet{Tremblay21a} showed that the Pearson linear correlation between the estimation and the answer was approximately 0.8.
The correlation increases when the horizontal velocity fields are averaged over several granular lifetimes \citep{Tremblay18};
the increment of the accuracy by taking the average shows the same trend as the LCT.
The results of DeepVel and LCT are similar when they are averaged,
however, DeepVel still has the advantage of reproducing the kinetic power spectra on sub-supergranule scales.
\citet{Tremblay20} developed a new architecture for DeepVel using the U-NET architecture
and found that it is more effective than other tracking methods.
However, their accuracies were not verified at various spatial scales.

There are several concerns of estimating the horizontal velocity on small scales,
which should be clarified before we use the method for observational data.
The motivation of this study was to evaluate the accuracy of the above methods
on various spatial scales
and reveal their weaknesses and possible improvements.
We developed a new method for estimating the spatial distribution of the horizontal velocity
based on a multi-scale deep learning architecture with several sizes of convolutional kernels
to capture the multi-scale nature of the solar convection.
We adopted the new method to three different numerical simulations of convection
and discussed the relationship between the power spectra of the velocities and performance of the network.
Furthermore, we suggest a new measure for evaluating the scale-by-scale velocity estimation.

\begin{figure}
  \centering
  \includegraphics[width=9cm]{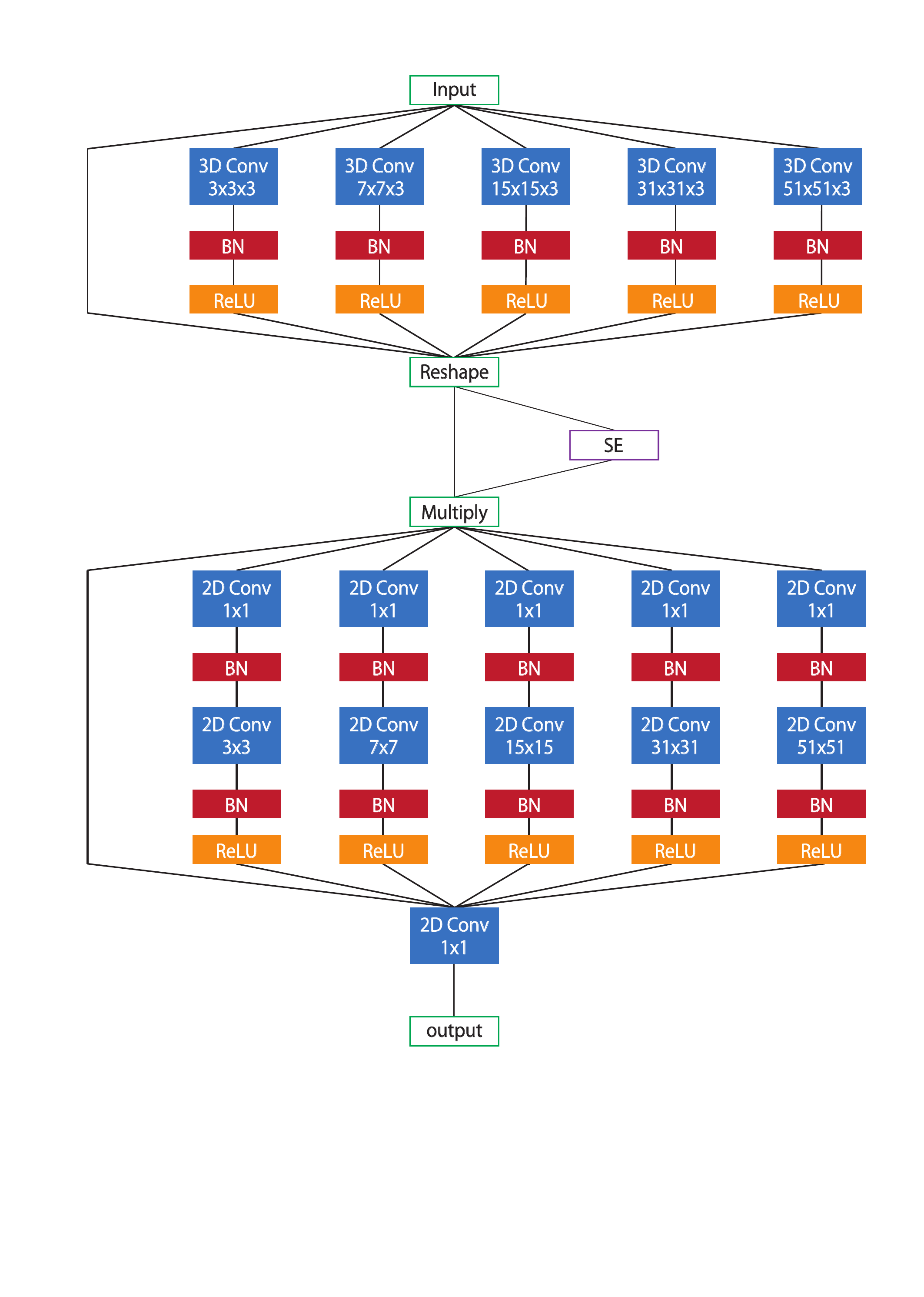}
  \caption{Structure of the network.
  BN and SE denote the batch normalization and squeeze-and-excitation, respectively.
  The number of kernels in each convolutional layer is presented in Table \ref{tb:kernels}.}
  \label{fig:network_structure}
\end{figure}

\begin{table}
\caption{Number of kernels in each convolutional layer.}
\centering
\begin{tabular}{l||ccccc}
          & $3\times3$ & $7\times7$ & $15\times15$ & $31\times31$ & $51\times51$\\
\hline\hline
3D Conv   & 60 & 40 & 20 & 10 & 5\\
\hline
2D Reduce & 20 & 10 &  5 &  5 & 2\\
\hline
2D Conv   & 60 & 40 & 20 & 10 & 5\\
\end{tabular}
\tablefoot{'2D Reduce' indicates $1\times1$ convolution before each 2D convolutional layer.}
\label{tb:kernels}
\end{table}

\begin{figure*}
  \centering
  \includegraphics[width=15cm]{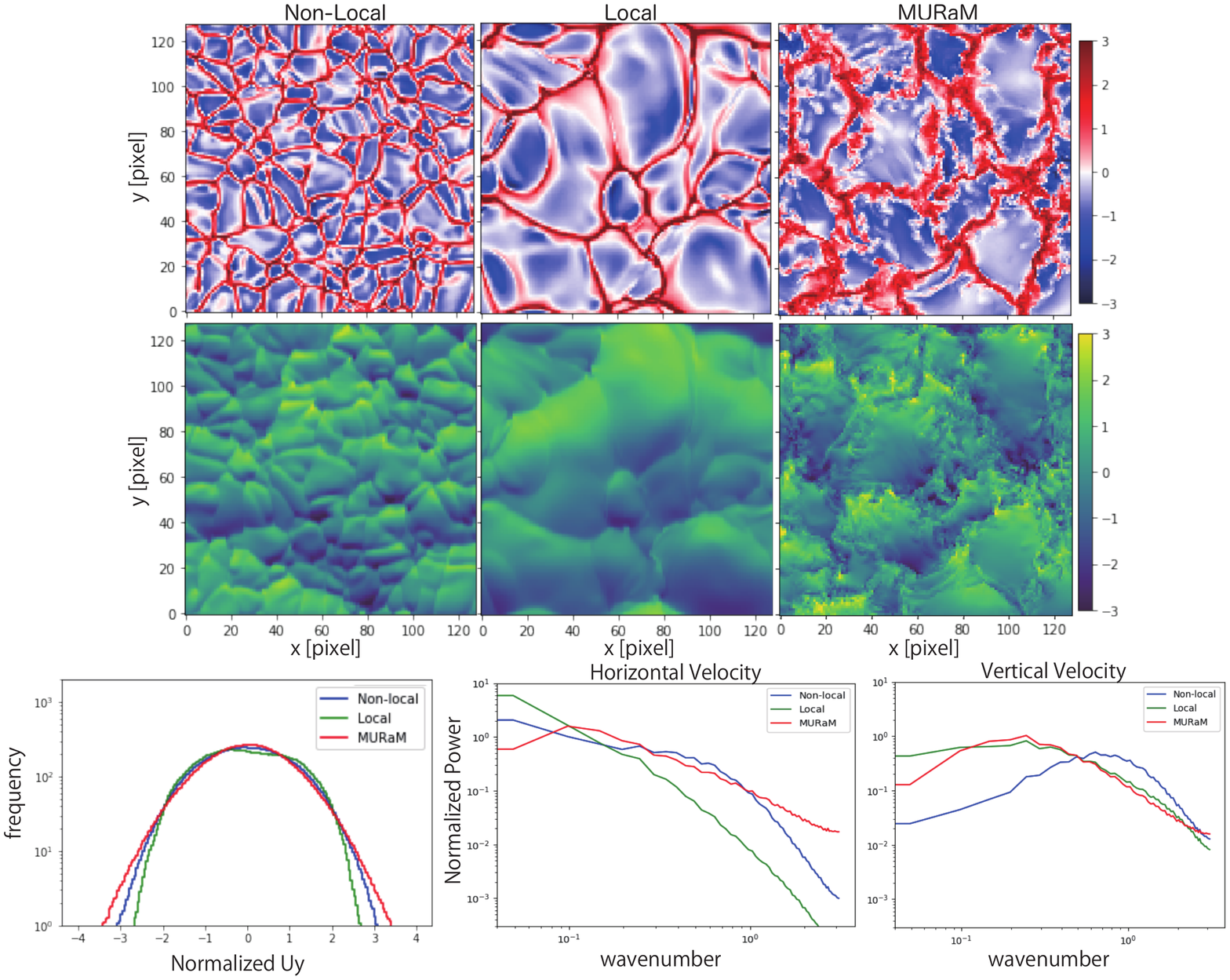}
  \caption{Representation of the training data sets.
  The spatial distributions of the vertical velocities (top panels) and
  Y component of the horizontal velocities (middle panels) of
  the non-local (left), local (center), and MURaM (right) simulation data.
  The bottom panels show the probability density functions (left),
  power spectra of the horizontal velocities (center),
  and power spectra of the vertical velocities (right) of the three simulations.
  The vertical and horizontal velocities in the simulations are normalized with zero and unit dispersion. This
  satisfies Equation (\ref{eq:ps_integration}).}
  \label{fig:data_sample}
\end{figure*}

\section{Method}
We developed a convolutional neural network that predicts the spatial distribution of the horizontal velocity
from the spatial and temporal variations of vertical velocity and temperature\footnote{
The trained network and sample data set can be found in \url{https://github.com/RT-Ishikawa/MultiScaleDL}.}.
This model includes multi-scale deep learning architectures:
the convolution layers have various sizes of the receptive fields (Figure \ref{fig:network_structure} and Table \ref{tb:kernels}).
The sizes of the kernels corresponded to 3$\times$3, 7$\times$7, 15$\times$15, 31$\times$31, and 51$\times$51.
This type of multi-scale architecture exhibits an advantage in detecting the solar convection motion, which is 
highly turbulent to the extent that the horizontal velocities exhibit broad power spectra.
This architecture is similar to the inception module \citep{Szegedy14}.
The inception module is a network that consists of kernels with various sizes, i.e., 1$\times$1, 3$\times$3, and 5$\times$5, and the pooling layer.
By utilizing the kernels with varying sizes, the inception module can not only efficiently detect spatially concentrated structures in a single region but also highly spread structures.

In the first block of the model, 3-dimensional convolutions were conducted along the spatial and temporal axes, and the channels corresponded to the different physical quantities.
After the first block, the data was reshaped, and thereby time and physical quantities were converged into the new channels.
Then, in the second block, the convolutions were conducted only along the spatial axes.
A potential problem corresponds to the large number of parameters due to the large size of kernels corresponding to 31$\times$31 and 51$\times$51.
To reduce the number of parameters, we also included 1x1 convolutions before each convolution layer.
Furthermore, some of the outputs of the convolutions were highly correlated.
This can decrease the efficiency of the optimization. The 1$\times$1 bottle-neck layer can partially resolve this problem.
After the 3-dimensional convolutions, the feature maps have 4 dimensions: two spatial sxes, temporal axes, and channels.
The reshape layer concatenate the temporal axes and the channels to change the structure of the feature maps into 3 dimensions.
The squeeze-and-excitation block \citep{Hu18} can improve the performance of the network
by modelling interdependencies between the channels.
This squeeze-and-excitation block produces a collection of modulation weights for the channels.
These weights are applied to the feature maps by multiplying and the results are fed into the subsequent layers.
Additionally, we included the batch normalization \citep{Ioffe15} after all the convolutions.
This normalized the outputs of convolutions into zero mean and unit covariance, and this in turn accelerated the training.
All the convolution layers were initialized with a random method \citep{He15a}.
\citet{Asensioramos17} developed a deep neural network model (DeepVel) to estimate the horizontal velocity.
Their model was based on ResNet \citep{He15b}, which consists of deeply stacked layers with only 2-dimensional convolutions
\footnote{\citet{Asensioramos17} that describe the convolution as ‘3-dimensional’.
The convolution was performed along spatial axes (x and y axes) and also dealt with the channels.
In this paper, we termed this as ‘2-dimensional’ after the procedure name ‘Conv2D’ in Keras. 
The 3-dimensional convolution in our network represents the convolution along the spatial and time axes as channels.}
with a receptive field of 3$\times$3.
The total number of trainable parameters of our model corresponded to $\sim 4.0\times10^5$, which was less than that of DeepVel by a factor of 4.


\begin{figure*}[tbp]
  \centering
  \includegraphics[width=18cm]{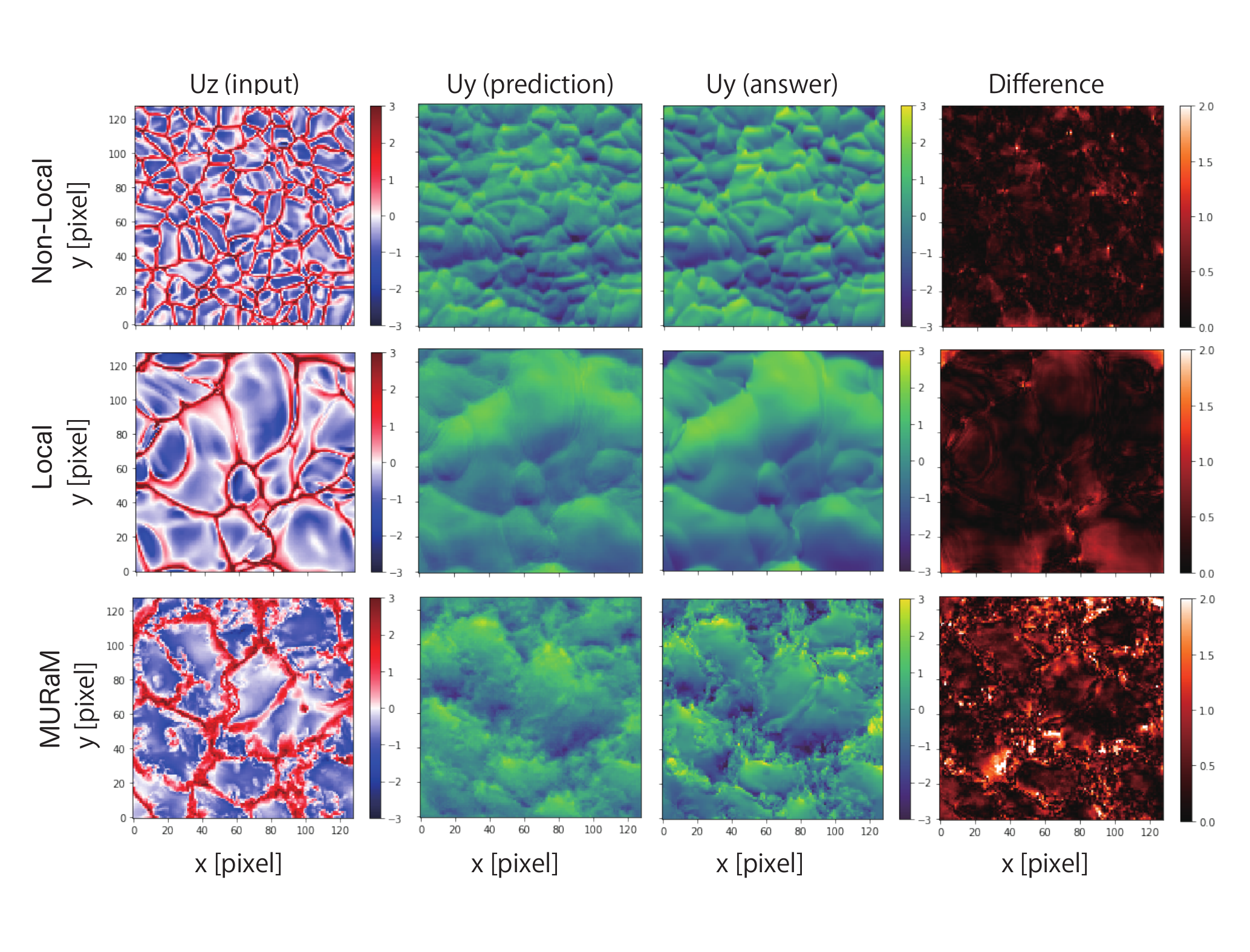}
  \caption{Spatial distributions of the vertical velocity, Y component of the horizontal velocities during the prediction and simulation,
and the difference between them
in the Non-local model (top), the Local model (middle), and the MURaM model (bottom), respectively.
The vertical velocity and horizontal velocity in the simulation are normalized as zero mean with unit dispersion.
  It should be noted that the vertical velocity and temperature are used for the inputs even though only the vertical velocity is shown in this figure.}
  \label{fig:result_map}
\end{figure*}

\section{Data}
To train the neural network,
we used numerical simulation data in three different types of convection models.
By comparing several cases with different energy injection scales and spectral properties,
we can evaluate and discuss the versatility of the neural network.

The first model is a convection model, in which convection is driven via cooling at the top boundary.
We termed this model as {\it non-local} one.
The 3-dimensional compressible MHD turbulence without any rotations
was considered in a Cartesian domain by covering the depth of the convection zone \citep{Masada16}.
The horizontal sizes of the domain were four times larger than the vertical size.
The super-adiabatic condition with the superadiabaticity $\delta \simeq 10^{-5}$
was imposed on only 5\% of the region to ensure that the top boundary was convectively unstable,
whereas the remaining part of the region remained adiabatic to ensure convective stability.
The cool downward plumes produced near the top boundary drove the convection:
fast downward motions with the entrainment behavior appeared locally and transiently. 

The second model was termed as {\it local} model. This
is a convection model, in which convection is driven by local entropy-gradient.
The entire convective zone was convectively unstable with super-adiabatic conditions.
The same Cartesian domain was considered as that in the non-local model.
Unlike the non-local model, convective cells were generated over the entire convective zone,
in which the spatial scale of cells were dependent on the local scale height in the vertical direction \citep{Cossette16}.
Hence, the convective motions with larger cells, which were produced near the bottom region,
were more pronounced in the local model when compared to those in the non-local model.

In the numerical simulations of the non-local and local models,
a spatial resolution of $256\times 256\times 128$ was used.
The other physical and numerical parameters were similar to those in an earlier study by \citet{Masada16}
(see \citealt{Yokoi21} for details.)
The apparent convective cells in the non-local and local simulations
roughly correspond to supergranular cells rather than granular ones.
The spatio-temporal data of the simulated turbulent fields were utilized for training the network,
where the MHD turbulence was fully developed to the statistically steady state
after several tens of the turnover time of the convective cells.
We used the temperature and vertical velocities at 3.3 Mm below the top boundary as the input data of the network.
Furthermore, the horizontal velocities were at the same depth as the corresponding ground truth.
We performed downsampling for these data by setting their sizes to 128$\times$128 pixels, which
covered a region of 748 Mm$\times$748 Mm.
Specifically, 1000 snapshots were available for both the models.
Because the input data do not represent the parameters on the surface,
the trained network is not directly applicable to an observation.
Instead, we compare the characteristics of the network with those trained with a realistic simulation, as described below.

We also used a three-dimensional compressible radiation MHD simulation, which was calculated
with the MURaM code \citep{Vogler05}, as the third model.
In the code, the radiative energy exchange was solved
via a non-gray radiative transfer under the assumption of local thermal equilibrium
(LTE) to reproduce more realistic granular scale flows in the photosphere.
Hence, in this study, we used the same simulation setup as that used by \citet{Riethmuller14}, in which
the simulation box covered 6 Mm in both horizontal directions with a grid size
of 10.42 km and 1.4 Mm in the vertical direction with a 14-km grid size.
A unipolar homogeneous vertical magnetic field of $B_z$ = 30 G
was introduced as an initial
condition, and the simulation was run for a total solar time of 16 h, which was long enough to reach a statistically stationary state.
We obtained more than 1600 snapshots of the data
cube with a mean cadence of 35 s.
We extracted a small region, which covers 5.4 Mm in both the horizontal directions,
and downsampled the region with a grid size of 42 km. Thus, the
size of the region corresponded to 128$\times$128 pixels.
We used the temperature and vertical velocity at the surface with an optical depth of unity.

The snapshots of the three simulations are shown in Figure \ref{fig:data_sample}.
The top three panels show the spatial distributions of the vertical velocities ($U_z$),
while the middle three panels show those of the horizontal velocities ($U_y$).
The vertical and horizontal velocities of each model were normalized with averages and standard deviations.
The bottom left panel shows the probability density functions of the horizontal velocities of the three simulations.
The fast velocity component existed in certain localized regions within short lifetimes.
This appeared as a non-gaussian distribution in the probability density function.
The power spectra of the horizontal and vertical velocities are shown in the bottom center and right panels.
Here, the power spectrum of a physical quantity $X$ is defined as follows:
\begin{equation}
E_X(k) = \frac{1}{2\Delta k} \sum_{\sqrt{k_x^2+k_y^2}\in[k-\Delta{k}/2,k+\Delta{k}/2)} |\hat{X}(k_x,k_y)|^2,
\end{equation}
where $\Delta k$ denotes the sampling interval of the wavenumber
and $\hat{X}$ denotes the Fourier transform
of the spatial distributions of the physical quantity $X$.
The Fourier transform of $X$ is as follows:
\begin{equation}
\hat{X}(k_x, k_y) = \frac{1}{N^2}\sum_{x,y}X(x,y)\exp[-i(k_xx+k_yy)],
\end{equation}
where $N^2$ denotes the number of pixels in an image,
which corresponded to 128$\times$128.
Then, the power spectrum was consistent with that defined in a previous study \citep{Rieutord10}
and satisfied
\begin{equation}
\frac{1}{2}\overline{X^2} = \int_0^{\infty}E_X(k) dk, \label{eq:ps_integration}
\end{equation}
where the overline denotes the average over the entire FOV.
Given the normalization of the vertical and horizontal velocities,
the dispersion $\overline{X^2}$ corresponded to unity.

The power spectra of the vertical velocity showed their peaks.
These peaks corresponded to the 'typical' scales of the convective cells,
while the spectra were broad.
The power spectra of the horizontal velocity exhibited high power on large-scales,
whereas they did not exhibit clear peaks,
especially in the non-local and local models.

\section{Training process}
We used the spatial distributions of temperature and vertical velocity
with a size of 128 $\times$ 128 pixels and three consecutive frames as the input data.
The output of the network was the spatial distribution of the horizontal velocity ($U_x$ or $U_y$).
We used the mean squared error between the horizontal velocities predicted by the network and those in the original data set
for the loss function. The loss function was expected to be minimized in the training.
Each physical quantity in the data set was normalized to zero average and unit standard deviation.
We prepared 350 sets of data for the training, 40 sets for the cross-validation, and 40 sets for the test.
Although the snapshots in the data sets were temporally consecutive,
the training, cross-validation, and test data sets
were separated sufficiently with an ample time interval that was
longer than the turn-over times of the convection in all the simulations.
The optimization of the network was performed by using
the Adam first-order gradient-based optimization \citep{Kingma14}.

Given the normalization of the training data set,
the average and standard deviation of the horizontal velocity predicted by the network
were also near zero and unity, respectively.
When we develop a network for application to an actual observational data,
the network should be supervised by realistic numerical simulations of the photosphere, such as MURaM,
with the spectral line synthesis and degradation with respect to the point-spread-function of observational instruments.
The network should predict the absolute value of the horizontal velocity without any normalization.
Hence, restoration of the velocity from the normalized value after the prediction
by multiplying the standard deviation of the horizontal velocity in the simulation
or that obtained with recent observations (e.g., \citealt{Oba20})
can potentially be an option.

\section{Results}
In this section, we show the comparison between the original simulation and model prediction of the horizontal velocities
for the three convection models.
In the LCT and DeepVel,
only the emergent intensity distributions were used for the estimation.
Furthermore, we used the vertical velocity distributions for the input of the network
and examined their benefits.
Table \ref{tb:corrcoef} lists the results with different inputs for each convection model.
The results showed that higher accuracy of estimation was realized by using the vertical velocity when compared to the temperature distributions. 
This result is consistent with \citet{Tremblay20}
By using vertical velocity and temperature, the network exhibited the highest global-correlation-coefficient.
However, it was slightly smaller than that obtained by using only vertical velocity for the local model.
In the rest of this paper, we focus on the network that uses the vertical velocity and temperature as the input.

\begin{table}
\caption{Global-correlation-coefficients with different inputs.}
\centering
\begin{tabular}{l||ccc}
          & Non-local & Local & MURaM\\
\hline\hline
$U_z$, $T$   & 0.948 & 0.948 & 0.771 \\
\hline
$U_z$      & 0.942 & 0.950 & 0.738 \\
\hline
$T$          & 0.890 & 0.884 & 0.668
\end{tabular}
\label{tb:corrcoef}
\end{table}

\begin{figure}
  \centering
  \includegraphics[width=9cm]{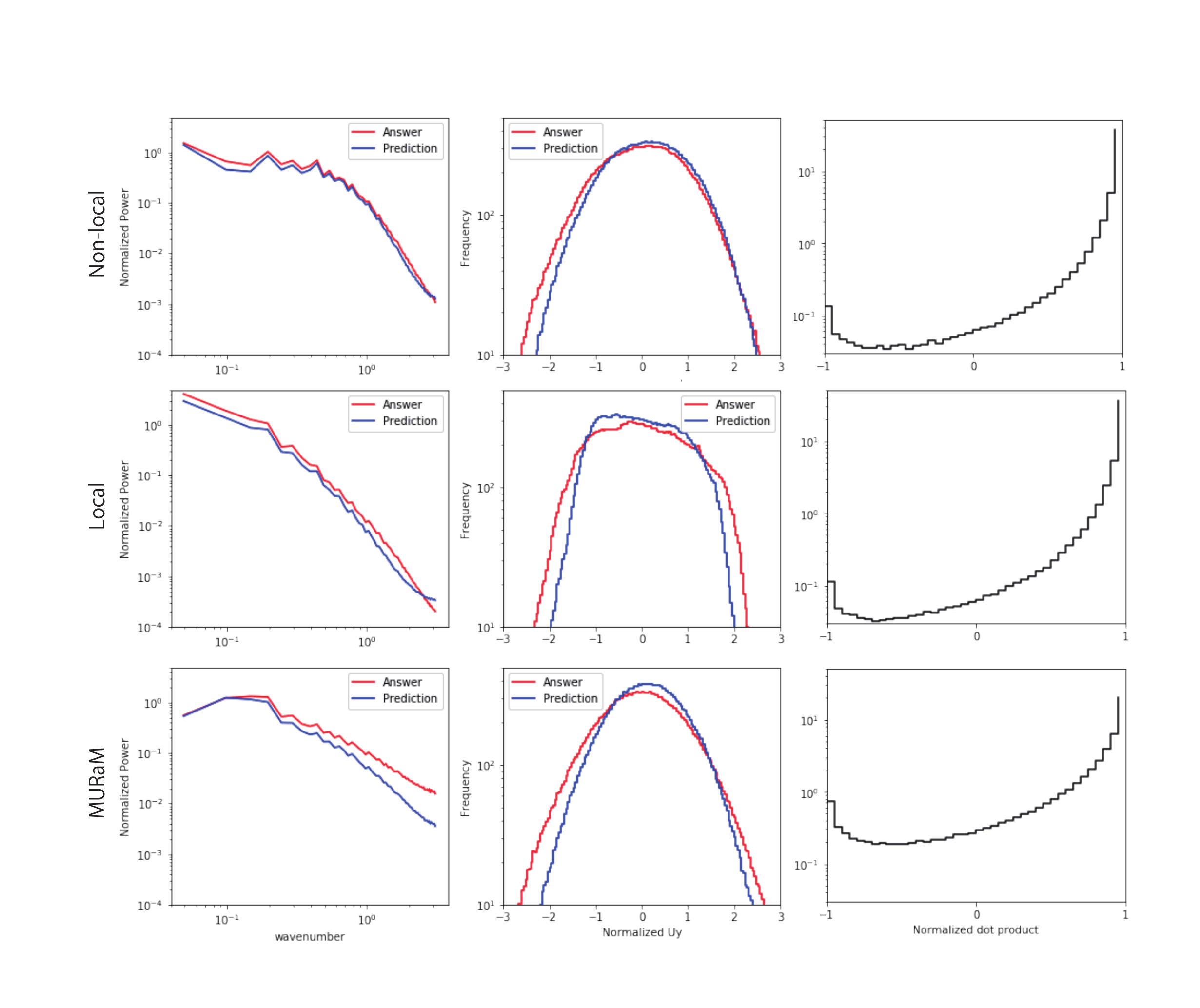}
  \caption{Comparison between
the simulated velocity field (red) and predicted velocity (blue)
in terms of power spectra (left) and histograms (center).
The histograms of normalized dot product (see Eq.\ref{eq:dot_product}) are shown in the right column.}
  \label{fig:result_powerspectra}
\end{figure}

\begin{figure}
  \centering
  \includegraphics[width=9cm]{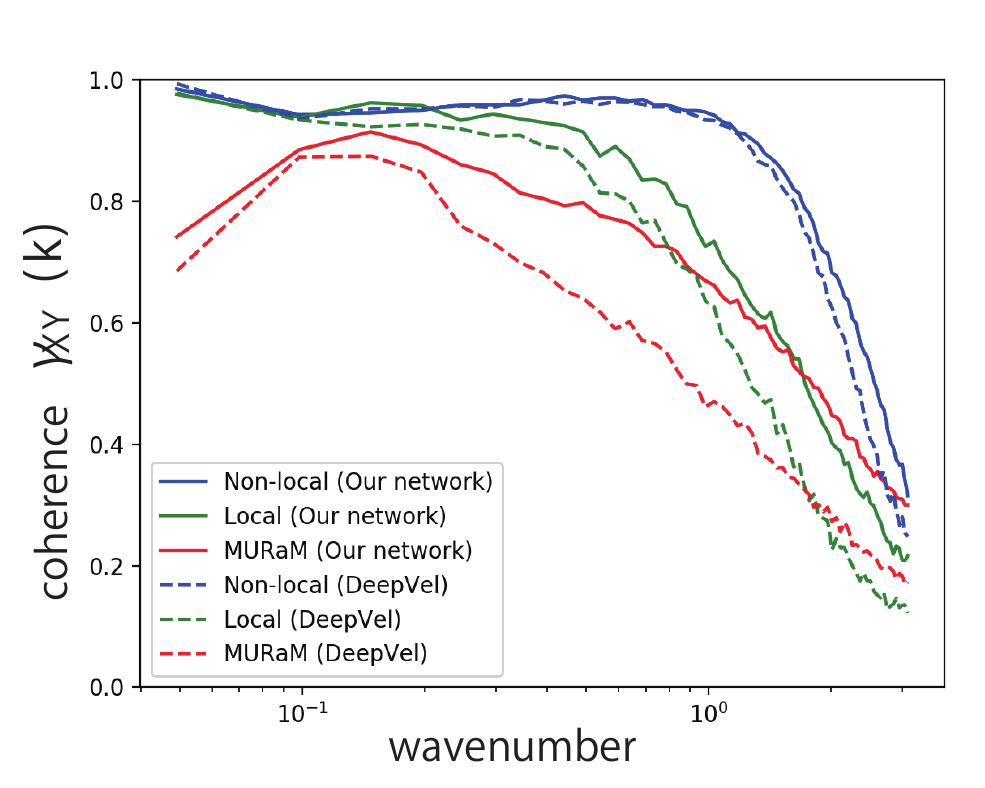}
  \caption{Coherence spectra between the simulation and prediction
as a function of the wavenumber with our network (solid) and DeepVel (dashed; \citealt{Asensioramos17}).
It should be noted that the vertical velocity distribution as opposed to the emergent intensity is used for the input to DeepVel .
The coherence spectra show strong scale-dependence.}
  \label{fig:result_correlate}
\end{figure}

Figure \ref{fig:result_map} compares the horizontal velocity distributions in the simulation and prediction
for each convection model.
In the figure, similar cellular patterns are observed in the simulated and predicted distributions.
The global-correlation-coefficients were 0.948, 0.948, and 0.771 for the non-local, local, and MURaM models, respectively.
The performance of the network for the MURaM model was worse when compared with those for the non-local and local models
in terms of the global-correlation-coefficients, and the predicted patterns appeared blurred.
The right column in Figure \ref{fig:result_map} shows the differences between the simulations and predictions.
The difference was expected to be likely high at the boundaries between the convection cells.
This was similar to the trend observed in the case of the LCT method \citep{Louis15}.
Although only the results for the $U_y$ are shown in the Figure,
the results for the $U_x$ are similar to those
because of the symmetric setup of the simulation and the network.

Figure \ref{fig:result_powerspectra} shows the comparisons between
the simulated velocity fields and predicted velocity fields in terms of the power spectra and histograms.
The power spectra of the predicted horizontal velocity distributions were similar to those of the simulation in the larger scales.
However, the power spectra were slightly underestimated in all the scales and noticeable errors were observed in the smaller scales.
In the case of the MURaM model, the network significantly underestimated the velocity in small-scales.
The occurrence of the fast velocities became less frequent in the predictions.
The blur in the images blended the fast velocity components which were preferentially localized in the small regions.

We also investigated dot products between the estimated and simulated horizontal velocity vectors.
The normalized dot product at each pixel is defined as follows:
\begin{equation}
A = \frac{\bm{v}\cdot\bm{v}_0}{||\bm{v}||~||\bm{v}_0||}, \label{eq:dot_product}
\end{equation}
where $\bm{v}$ and $\bm{v}_0$ represent the estimated and simulated horizontal velocity vectors, respectively.
This definition is consistent with \citet{Tremblay20} and \citet{Tremblay21b}.
This dot product indicates the difference of the orientation of the vectors.
The histograms shown in the right column of Figure \ref{fig:result_powerspectra} have a peak around $A\sim1$,
which indicates that the estimated and simulated horizontal velocity vectors are well aligned in most regions.
The histograms have tails toward the negative dot products,
although the fraction of the tails is small.
The tail is relatively large in the histograms of the MURaM model.
The averages of the normalized dot products are
0.92, 0.91, and 0.70
for the non-local, local, and MURaM models, respectively.

To examine the performance of the network,
we defined the coherence spectrum, which represents the correlation at each wavenumber.
This has never been introduced in past studies including as \citet{Louis15} and \citet{Asensioramos17}.
The coherence spectrum of distributions of two physical quantities $X$ and $Y$ can be defined as follows.
First, we calculate the cross-spectrum of $X$ and $Y$ as follows:
\begin{equation}
S_{XY}(k) = \frac{1}{2\Delta k} \sum_{\sqrt{k_x^2+k_y^2}\in[k-\Delta k/2,k+\Delta k/2)} \langle\hat{X}^{*}\hat{Y}\rangle(k_x,k_y),
\end{equation}
where asterisk denotes the complex conjugate,
and the ensemble average $\langle\rangle$ is calculated over all the 40 frames in the test data.
Finally, we defined the coherence spectrum by normalizing the cross-spectrum with the average power spectra as follows:
\begin{equation}
\gamma_{XY}(k) = \frac{{S_{XY}}(k)}{\sqrt{\langle{E_X}\rangle(k)\langle{E_Y}\rangle(k)}}.
\end{equation}
The cross-spectrum $\hat{X}^{*}\hat{Y}$ is equivalent to the Fourier transform of the cross-correlation function of $X$ and $Y$.
Therefore, the coherence spectrum $\gamma_{XY}(k)$ represents the correlation between the two distributions at each wavenumber.
A useful relationship exists among the coherence spectrum, power spectra, and global-correlation-coefficient as follows:
\begin{equation}
R_{XY}(0,0) = \frac{\sum_{k}{\sqrt{\langle{E_X}\rangle\langle{E_Y}\rangle}\gamma_{XY}}(k)}{\sqrt{\sum_k\langle{E_X}\rangle(k)\sum_{k^{\prime}}\langle{E_Y}\rangle(k^{\prime})}}
\approx \frac{\sum_{k}{\langle{E_X}\rangle(k)\gamma_{XY}(k)}}{\sum_k\langle{E_X}\rangle(k)}, \label{eq:R_gamma}
\end{equation}
where $R_{XY}(0,0)$ denotes the global-correlation-coefficient without any translation (see Appendix \ref{sec:Appendix_A}).
This equation indicates that the global-correlation-coefficient is equivalent to
the average of the coherence spectrum weighted with the power spectra. Hence, this highlights the importance of the coherence spectrum analysis, shown in this study,  
in examining the prediction accuracy scale-by-scale,
particularly in small-scale structures.
As evident in the local model, the power spectrum of the horizontal velocity exhibited high power in large-scales.
Subsequently, the coherence spectra of large-scales exhibited high importance for the global-correlation-coefficient.
Conversely, the power spectrum of the horizontal velocity in the MURaM model exhibited
relatively high power in small-scales when compared to the non-local and local models.
This indicated higher importance of the coherence spectra of small-scales.

The results of the coherence spectra $\gamma_{XY}(k)$ for the three convection models are shown in Figure \ref{fig:result_correlate}.
We reconstructed the DeepVel and trained it with three convection simulations with vertical velocity as the input,
while the original DeepVel was trained with the emergent intensity.
The performance of our network was similar or better than that of DeepVel for all the simulation data and at all the wavenumbers with the exception of the large-scale in the non-local model.
However, it was still difficult to reproduce the small-scale structures.
In the non-local and local models,
the coherence spectra exceeded 0.9 for small wavenumbers
and they drastically decreased to 0.3 or less for large wavenumbers.
Furthermore, even the global-correlation-coefficients for both the models exceeded 0.9.
The discussion on the small-scale structures should be treated with care even if the global-correlation-coefficient is high.
In the MURaM model, which exhibited lower steep power spectra of the horizontal velocity (Figure \ref{fig:data_sample}),
the performance was limited in a wide range of wavenumbers when compared with those of the non-local and local cases.
However, a rough trend and better accuracy in large-scale were observed.

When comparing the results of the three convection models,
it can be observed that the different onset of the significant decrease
in the coherence spectrum appears
on smaller scales as opposed to on energy injection scales
where the power spectra of the vertical velocities have their peaks.
This signature suggests that the decrease in performance is not
simply due to the small size of the spatial patterns, but also due to a reason related to turbulent structures.
For example, the energy injection scale of $k\sim0.7$,
which is observed from the power spectrum of the vertical velocity in the non-local case (Figure \ref{fig:data_sample}),
is correlated to the onset wavenumber for degrading the prediction accuracy toward the high wavenumber region.
This is attributed to the non-linear cascade of convective eddies.
As presented in Table \ref{tb:corrcoef},
the high performance of this network relies on the vertical velocity.
This implies that the relationship between the horizontal and vertical flows is rather simple in large-scales,
whereas it is more complex in small-scales probably due to the non-linearity.

The small global-correlation-coefficient for the MURaM model
is related two characteristics of the convection:
large energy injection scale
and small slope of the power spectra.
The energy injection scale is higher than that of the non-local model,
which diminishes the coherence spectra in a wide range of spatial scales.
In combination with the higher power in the small-scales,
the resultant global-correlation-coefficient becomes small for the MURaM model.
A large global-correlation-coefficient was realized for the local model
even though the energy injection scale was similar to that in the MURaM model.
This was due to the fact that the power spectrum of the horizontal velocity in the local model exhibited a steep slope, which in turn
induced large weights on the large-scales.

To improve the accuracy at smaller scales,
the optimization of the small kernels should be performed more efficiently.
A potential method of realizing this involves improving the loss function.
In this study, the network was trained by minimizing the mean-squared-error
between the prediction and simulation.
Given that the horizontal velocity distribution exhibits high power in large-scales,
as shown in Figure \ref{fig:data_sample},
larger weights are used on the larger scale structures for the loss function.
This can promote the learning of the large-scale structures
and can inhibit the detection of the small-scale structures.
Hence, an appropriate loss function can improve the optimization.
An alternative method involves increasing the number of physical quantities to the input.
In this study, the spatial distributions of the vertical velocity and temperature
were used as the input data.
It can be beneficial to use other physical quantities, which are associated with the non-linear process
such as the velocities at multiple heights.
Furthermore, the spatial distributions with small-scale enhancement obtained via pre-processing
can also aid in improving the performance.

\section{Conclusions}
We developed a novel convolutional neural network to estimate the spatial distribution of horizontal velocity
by using the spatial distributions of temperature and vertical velocity. 
This new network was comprised of convolutional layers with various sizes of receptive fields.
This led to efficient detection of spatially spread features and concentrated features.
Given that the velocity distribution driven by the convection exhibited multi-scale structures and broad power spectra,
this type of a multi-scale network exhibited an advantage in detecting the solar surface convective motion.
Our network exhibited a higher performance on almost all the spatial scales when compared to those reported in previous studies.

In most previous studies, the evaluation of the method was performed
solely via the global-correlation-coefficient
between the simulated velocity field and prediction velocity field.
This in turn cannot show the accuracy on different spatial scales.
To evaluate the accuracy at each scale,
we introduced the coherence spectrum, which represents the correlation at each wavenumber.
The evaluation with the coherence spectrum revealed the strong scale-dependence of the accuracy.
The horizontal velocities at large-scales were well predicted with our network,
while those at small-scales were limited.
Recently, the small-scale vortical motions in the solar photosphere has attracted significant attention
observationally \citep{Bonet08, Vargas11} and theoretically \citep{Shelyag11,Moll11}.
However, given the rapid decrease of the accuracy in estimating the horizontal velocities
toward the small-scales,
we should be careful when discussing them observationally.
The accuracy for the small-scale structures is crucial
for calculating vorticities via a derivative of horizontal velocities.

By comparing the results of the three convection models,
we observed that the rapid decrease in coherence spectrum occurred on the scales that were lower than the energy injection scales, which were characterized by the peaks of the power spectra of the vertical velocities.
This implies that the network was not appropriately trained
to reproduce the velocity fields in small-scales generated by turbulent cascades.
To improve the accuracy on small-scales,
it is potentially important to consider improving the loss function in the network. This can be realized by adding more weights on the small-scale structures
and inputting other physical quantities, such as the vertical velocities at multiple heights,
which can be related to the non-linear process.
These challenges can be explored in future studies.

In this study we developed a new model for estimating the horizontal velocity field
with numerical simulation data
and evaluated its performance with a new measure.
By adopting our network
to the high-resolution observation data
obtained via the SUNRISE-3 balloon-born telescope \citep{Katsukawa20,Feller20}
and Daniel K. Inouye Solar Telescope \citep{Rimmele20,Rast21},
we can estimate the horizontal velocity distributions
with a high accuracy in a wide range of wavenumbers.

\begin{acknowledgements}
R.T.I. is supported by JSPS Research Fellowships for Young Scientists.
This study is supported by the NINS program for cross-disciplinary study
on Turbulence, Transport, and Heating Dynamics
in Laboratory and Astrophysical Plasmas: “SoLaBo-X” (Grant Numbers 01321802 and 01311904)
and by JSPS KAKENHI Grant Numbers JP18H05234 (PI: Y. Katsukawa), JP19J20294 (PI: R.T. Ishikawa), and 20KK0072 (PI: S. Toriumi).
\end{acknowledgements}

%
%

\begin{appendix} 
\section{Coherence Spectrum Analysis}\label{sec:Appendix_A}
We introduce a new measure {\it coherence spectrum} for evaluating the network.
In this section, we describe the definition of the coherence spectrum
and provide a proof of Equation (\ref{eq:R_gamma}).
When we have the spatial distributions of physical quantities $X$ and $Y$,
we can define the 2-dimensional cross-spectrum by using the Fourier transform as follows:
\begin{align}
S_{XY}(k_x,k_y) &= \hat{X}^{*}\hat{Y}, \\
\hat{X}(k_x,k_y) &= \frac{1}{N^2}\sum_{x,y}X(x,y)\exp[-i(k_xx+k_yy)].
\end{align}
The power spectrum is as follows:
\begin{equation}
E_X(k) = \frac{1}{2\Delta k} \sum_{\sqrt{k_x^2+k_y^2}\in[k-\Delta{k}/2,k+\Delta{k}/2)} |\hat{X}(k_x,k_y)|^2,
\end{equation}
where $\Delta k$ is the sampling interval of the wavenumber.
Furthermore, the cross-spectrum is equivalent to the Fourier transform of the correlation function between $X$ and $Y$,
which corresponds to the Wiener--Khinchin theorem.
\begin{equation}
C_{XY}(\xi,\eta) = \sum_{k_x,k_y}S_{XY}(k_x,k_y)\exp[i(k_x\xi+k_y\eta)].
\end{equation}
Then, we obtain the relationship between the global-correlation-coefficient $R_{XY}$
and 2-dimensional cross-spectrum:
\begin{equation}
R_{XY}(0,0) = \frac{1}{\sqrt{C_{XX}(0,0)C_{YY}(0,0)}}\sum_{k_x,k_y}S_{XY}(k_x,k_y).
\end{equation}
Here, we define the cross-spectrum by calculating the ensemble average $\langle\rangle$
and integrating the 2-dimensional cross-spectrum as follows:
\begin{equation}
S_{XY}(k) = \frac{1}{2\Delta k}\sum_{\sqrt{k_x^2+k_y^2}\in[k-\Delta k/2,k+\Delta k/2)}\langle S_{XY}\rangle(k_x,k_y).
\end{equation}
The phase of the averaged 2-dimensional cross-spectrum corresponds to the 'average' phase difference
between the two Fourier components.
Given the contour integration, the imaginary part of the cross-spectrum converges to zero
because $S_{XY}(-k_x,-k_y)=S_{XY}^{*}(k_x,k_y)$ is always satisfied.
This integration is justified when the statistical isotropy is assumed or else the resultant coherence spectrum can be misleading.
Consequently, the coherence spectrum defined below is a real number.
\begin{equation}
\gamma_{XY}(k) = \frac{{S_{XY}}(k)}{\sqrt{\langle{E_X}\rangle(k)\langle{E_Y}\rangle(k)}},
\end{equation}
Hence, we obtain the following.
\begin{equation}
R_{XY}(0,0) = \frac{\sum_{k}{\sqrt{\langle{E_X}\rangle\langle{E_Y}\rangle}\gamma_{XY}}(k)}{\sqrt{\sum_k\langle{E_X\rangle(k)}\sum_{k^{\prime}}\langle{E_Y\rangle(k^{\prime})}}},
\end{equation}
where we use equation $C_{XX}(0,0)=\overline{X^2}=2\sum_{k}E_{X}(k)\Delta{k}$
and condition of the statistical steady state $\overline{X^2}=\langle\overline{X^2}\rangle$.

\end{appendix}

\end{document}